\documentclass[preprint2]{aastex}
\usepackage{graphicx}
\usepackage{amsmath}
\usepackage{amsfonts}
\usepackage{amssymb}
\usepackage{epsfig}
\begin{document}

\title{Constraining the Quintessence equation of state \\
with SnIa data and CMB peaks}

\author{P.S. Corasaniti\altaffilmark{1} and E.J. Copeland\altaffilmark{2}}
\affil{Centre for Theoretical Physics, University of Sussex, Brighton, 
BN1 9QJ\\United Kingdom}
\altaffiltext{1}{e-mail: pierc@pact.cpes.susx.ac.uk}
\altaffiltext{2}{e-mail: e.j.copeland@sussex.ac.uk}

\begin{abstract}
Quintessence has been introduced as
an alternative to the cosmological constant scenario to account
for the current acceleration of the universe.
This new dark energy component allows values of the equation of 
state parameter
$w_{Q}^0\geq-1$, and in principle measurements of cosmological distances 
to Type Ia supernovae can be used
to distinguish between these two types of models. 
Assuming a flat universe, we use the supernovae data
and measurements of the position of the acoustic peaks in the 
Cosmic Microwave Background (CMB) spectra to constrain a 
rather general class of 
Quintessence potentials, including inverse power law models and recently 
proposed Supergravity inspired potentials. In particular we use a likelihood analysis, marginalizing over
the dark energy density $\Omega_{Q}$, the physical baryon density $\Omega_{b}h^2$
and the scalar spectral index $n$, to constrain
the slopes of our Quintessence potential. Considering only the first Doppler peak
the best fit in our range of models gives $w_{Q}^0\sim-0.8$. However, including
the SnIa data and the three peaks, we find an upper limit on the present value of
 the equation of state parameter, $-1\leq w_{Q}^0\leq-0.93$ 
at $2\sigma$, a result that appears to rule out a class of recently proposed 
potentials. 

\end{abstract}


\section{Introduction}

Observations of distant type Ia supernovae (Perlmutter et al. 1999;
Riess et al. 1999) and small angular scale anisotropies in the 
Cosmic Microwave
Background (CMB) (De Bernardis et al. 2000; Balbi et al. 2000;
Netterfield et al. 2001; Pryke et al. 2001) suggest that the 
universe is dominated 
by a
large amount of dark energy with a negative equation of state parameter
$w$. One obvious explanation would be the presence for all time of a 
cosmological constant with $w=-1$, although there is no satisfactory reason known why 
it should be so close to the  critical energy density
 (for a general review see Sahni \&\ Starobinsky 2000). An alternative 
 proposal
introduces a new type of matter and is called 'Quintessence' 
(Caldwell et al. 1998).
Assuming that some unknown mechanism cancels the true cosmological
constant, this dark energy is associated with a light scalar field
$Q$ evolving in a potential $V(Q)$. 
The equation of state parameter of the $Q$ component is given by
\begin{equation}
w_{Q}=\frac{\frac{\dot{Q}^2}{2}-V(Q)}{\frac{\dot{Q}^2}{2}+V(Q)}
\label{eqstate}
\end{equation} 
and it is a function
of time. According to the form of
$V(Q)$ the present value of $w_{Q}$ is in the range $w_{Q}^0\geq-1$.
 The temporal dependence 
of $w_{Q}$ implies 
that 
high red-shift observations could in principle distinguish between $\Lambda
CDM$ and $QCDM$ models (Maor et al. 2000; Huterer \&\ Turner 2000; Alcaniz \&\ Lima 2001; Benabed \&\ Bernardeau 2001; 
Cappi 2001; Weller \&\ Albrecht 2001).
Moreover, a number of authors have recently pointed out 
that
the position of the CMB peaks could provide an efficient way to constrain 
Quintessence
models (Kamionkowski \&\ Buchalter 2000; Croocks et al. 2000; Doran et al. 2000).

In this paper we use the supernovae sample of Perlmutter \textit{et al.} (1999) and the recent measurements 
of
the location of the CMB peaks  (De Bernardis et al. 2001) to determine new
limits on the Quintessence equation of
state. Our study is similar in approach to an earlier analysis 
by Efstathiou (2000).
We consider a general class of potentials parametrized in such a 
way that we can control their
shape, and apply a likelihood analysis to find the confidence
regions for the parameters of the potential and the best value for the fractional Quintessence energy 
density $\Omega_{Q}$. The constraints which emerge are different if we analyze the data separately.
In particular the position of the first Doppler peak prefers a Quintessence model with $w_{Q}^0\sim-0.8$
 for the prior $\Omega_{Q}=0.7$ in agreement with Baccigaluppi \textit{et al.} (2001),
while the analysis of all the CMB peaks and SnIa gives an 
upper value for the equation of state, $w_{Q}^0\leq-0.93$ at $2\sigma$ for these class of models. This limit is 
stronger than those previously obtained (Perlmutter et al. 1999; Efstathiou 2000; Amendola 2001; Balbi et al. 2001), 
$w_{Q}^0\leq-0.6$ at $2\sigma$, 
simply because we are making use of the new improved CMB data. An obvious consequence of this
result is that in these class 
of models, for them to succeed 
the scalar field dynamics has to produce
effects similar to pure vacuum energy and in this case it is unlikely that 
Quintessence can be distinguished from a cosmological constant (see also Maor et al. 2000).

\section{Quintessence equation of state}

The scalar field dynamics is described by the Klein-Gordon equation
\begin{equation}
\ddot{Q}+3H\dot{Q}+\frac{dV}{dQ}=0,
\label{klein}
\end{equation}
with 
\begin{equation}
H^2=\frac{8\pi G}{3}\left[\rho_{m}+\rho_{r}+\frac{\dot{Q}^2}{2}+V(Q)\right],
\label{friedmann}
\end{equation}
where $\rho_{m}$ and $\rho_{r}$ are the matter and radiation energy densities
respectively. It is well known that for a wide class of potentials, Eq.~(\ref{klein}) possesses
attractor solutions (Steinhardt et al. 1999). In this regime the 
kinetic energy of
the field is subdominant allowing $w_{Q}$ to become negative. The present value of
$w_{Q}^0$ depends on the
slope of the potential in the region reached by the field.
Actually if the
Quintessence field rolls down a very flat region (Barreiro et al. 2000) or if it evolves close to a minimum (Albrecht \&
Skordis 1999, Brax \& Martin 1999, Copeland et al. 2000) the equation of state parameter varies in the range
 $-1\leq w_{Q}^0<-0.8$. On the other hand models like the
inverse power law potential (Ratra \& Peebles 1988; Zlatev et al. 1999) require larger values of $w_{Q}^0$.
A general potential which can accomodate a large class of scenarios is:

\begin{equation}
V(Q)=\frac{M^{4+\alpha}}{Q^{\alpha}}e^{\frac{1}{2}(\kappa Q)^{\beta}},
\label{potenziale}
\end{equation}
where $\kappa =\sqrt{8\pi G}$ and $M$ is fixed in such a way that today $\rho
_{Q}=\rho_{c}\Omega_{Q}$, where $\rho_{c}$ is the critical energy density. For $\beta=0$ Eq.~(\ref{potenziale})
becomes an inverse power law, while for $\beta=2$ we have the 
SUGRA potential proposed by (Brax \&\
Martin 1999). For $\alpha=0$, $\beta=1$ and starting with a large value of $Q$,
the Quintessence field evolves in a pure exponential potential (Ferreira \&\ Joyce 1998).
We do not consider this case further since it is possible
to have a dark energy dominated universe, but at the expense of fine tuning for
 the initial conditions of the
scalar field. Larger values of $\beta$ mimic the model studied by Copeland
\textit{et al.} (2000). For $\alpha,\beta\neq0$ the potential has a minimum, the dynamics can
be summarized as the following. For small values of $\beta$ and for a large range of initial conditions,
 the field does not reach the minimum by the present time and hence $w_{Q}^0>-1$. For example,
if the Quintessence energy density initially dominates over the radiation, the $Q$ field quickly rolls down
 the inverse power law part of the potential eventually
  resting in the minimum with $w_{Q}\sim-1$ after a series of damped oscillations (Riazuelo \&\ Uzan 2000).
This behaviour however requires fine tuning the initial value of $Q$ to be small. On the other hand, this can be avoided
if we consider large 
values of $\alpha$ and $\beta$ (Fig.1a). In these models the fractional energy density of the
Quintessence field, 
$\Omega_{Q}$, is always negligible during both radiation
and matter
dominated eras. In fact, for small initial values of $Q$, $V(Q)$ acts like an inverse 
power law potential,
hence as $Q$ enters the scaling regime its energy density is subdominant 
compared to that of
the background component. Therefore 
nucleosynthesis
constraints (Bean et al. 2001) are always satisfied and there are no
physical effects on the evolution of the density perturbations. The main
consequence is that for a different value of $w_{Q}^0$ the
Universe starts to accelerate at a different red-shift (Fig.1b).
\begin{figure}[ht]
\plotone{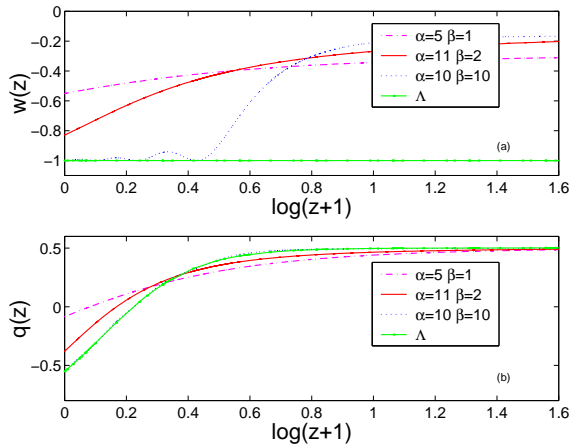}
\caption{In (a) the evolution of $w_{Q}$ against the red-shift  is plotted 
for different values of
$\alpha$ and $\beta$. In (b) the behaviour of the deceleration parameter, 
$q$, is plotted against the red-shift. The acceleration starts ($q<0$) earlier 
for models with an equation of state close to that of a true cosmological constant. 
\label{fig1}}
\end{figure}
This implies that  different values of $\alpha$ and $\beta$ lead to a 
different luminosity distance and angular diameter distance.
Consequently by making use of the observed distances we may in principle 
determine an upper limit on $w_{Q}^0$, potentially constraining the allowed shape 
of the Quintessence potential (Huterer \&\ Turner 2000).
\section{CMB peaks}

The CMB power spectrum provides information on combinations of the fundamental cosmological
quantities. The position of the Doppler peaks depends on the
geometry of the Universe
through the angular diameter distance, although the amplitude 
of the peaks are 
sensitive to many different 
parameters. The important point for us is that in general the  
Quintessence field can contribute to the shape of the 
spectrum through both the early 
integrated Sachs-Wolfe effect (ISW) and the late one
 (Hu et al. 1997). 
The former 
is important if
the dark energy contribution at the last scattering surface (LSS) 
is not negligible (Skordis \& Albrecht 2000, Barreiro et al. 2000) 
or in non-minimally coupled models (Amendola 2000, Perrotta et al. 2000, 
Baccigalupi et al. 2000), whereas
the late ISW is the only effect in models with $\Omega_{Q}\sim0$ at LSS 
(Brax et al. 2000).
However, as has recently been demonstrated an 
accurate determination of
the position of the Doppler
peaks is more sensitive to the actual amount of dark energy 
(Doran et al 2000).
To be more precise, the multipole of the $m$-th peak is 
$l_{m}=m l_{sh}$, where $l_{sh}$ is 
proportional
to the angular scale of the sound horizon at LSS.
In a flat universe $l_{sh}$ is given by:
\begin{equation}
l_{sh}=\frac{\pi}{\bar{c}_{s}}\left(  \frac{\tau_{0}}{\tau_{ls}}-1\right),
\end{equation}
where $\bar{c}_{s}$ is the mean sound velocity and $\tau_{0}$, $\tau_{ls}$ 
are
the conformal time today and at last scattering respectively.
However, physical effects before recombination can shift the scale of 
the
sound horizon at different multipoles, resulting in a 
better estimate for the peak 
positions being given by:
\begin{equation}
l_{m}=l_{sh}(m-\delta l-\delta l_{m}),
\label{peak}
\end{equation}
where $\delta l$ is an overall shift (W. Hu et al. 2000) and 
$\delta l_{m}$
is the shift of the $m$-th peak. These corrections depend on the 
amount of
baryons $\Omega_{b}h^{2}$, on the fractional quintessence energy density
at last scattering ($\Omega_{Q}^{ls}$) and today ($\Omega_{Q}^{0}$), as well as on the scalar 
spectral index $n$.
Recently, analytic formulae, valid over a 
large range of the cosmological parameters,
have been provided to good accuracy for 
$\delta l$ and $\delta l_{m}$ (Doran \&\ Lilley 2001).
Of crucial importance is the observation that
the position of the third peak appears to
remain insensitive to other
cosmological quantities, hence we can make use of this fact to test dark energy 
models (Doran et al. 2001).

\section{Likelihood analysis and results}

\subsection{Constraints from supernovae}

We want to constrain the set of parameters $\alpha$, $\beta$ and $\Omega_{Q}$
confined in the range: $\alpha\in(1,10)$, $\beta\in(0,10)$ and $\Omega_{Q}\in(0,1)$,
subject to the assumption of a flat universe.
We use the SnIa data fit C of Perlmutter \textit{et al.} (1999), that excludes 4 high redshift
data points. The magnitude-redshift relation is given by:
\begin{equation}
m(z)=\mathcal{M}+5\log\mathcal{D}_{L}(z,\alpha,\beta,\Omega_{Q}),
\end{equation}
where $\mathcal{M}$ is the `Hubble constant free' absolute magnitude and
 $\mathcal{D}_{L}=H_{0}d_{L}(z)$ is the free-Hubble constant luminosity
distance. In a flat universe
\begin{equation}
d_{L}(z)=(\tau_{0}-\tau(z))(1+z),
\label{dist}
\end{equation}
where $\tau_{0}$ is the conformal time today and $\tau(z)$ is the conformal
 time at the red-shift $z$ of the observed supernova. Both of these quantities are calculated
 solving numerically Eq.~(\ref{klein}) and Eq.~(\ref{friedmann}) for each value of $\alpha,\beta$ and $\Omega_{Q}$.
In $\mathcal{M}$ we neglect the dependence on a fifth paramenter
($\alpha$ in Perlmutter \textit{et al.} 1999) and assume it to be 0.6, the Perlmutter \textit{et al.} (1999) best value. 
We then obtain a gaussian likelihood
function $\mathcal{L}^{Sn}(\alpha,\beta,\Omega_{Q})$, by 
marginalizing over
$\mathcal{M}$. In Fig.2a we present the one-dimensional likelihood 
function normalized to
its maximum value for $\Omega_{Q}$. There is a maximum at $\Omega_{Q}=1$,
in agreement with Efstathiou (2000). In Fig.3a we present
the likelihood contours in the
$\alpha - \beta$ parameter space, 
obtained after marginalizing over $\Omega_{Q}$. 
Note that all values are allowed at the $2\sigma$ level.
The confidence regions for the SnIa data correspond to Quintessence models
with $w_{Q}^0<-0.4$ for $\Omega_{Q}=0.6$, an upper
limit that agrees with both Perlmutter \textit{et al.} (1999) 
and Efstathiou (2000).

\begin{figure}[ht]
\plotone{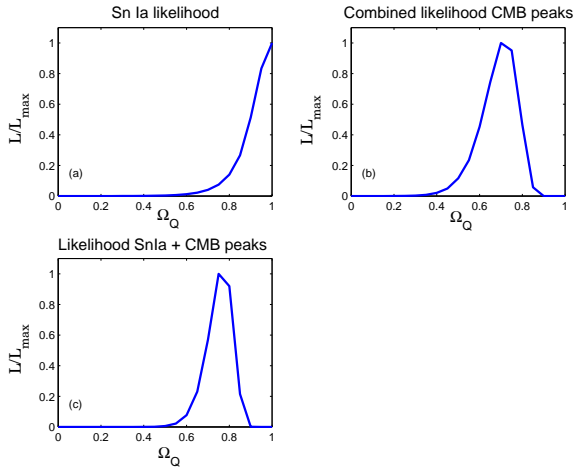}
\caption{ Fractional Quintessence energy density likelihoods, (a) for SnIa,
(b) for the combined CMB peaks and (c) for the combined 
data sets.\label{fig2}}
\end{figure}
\begin{figure}[ht]
\plotone{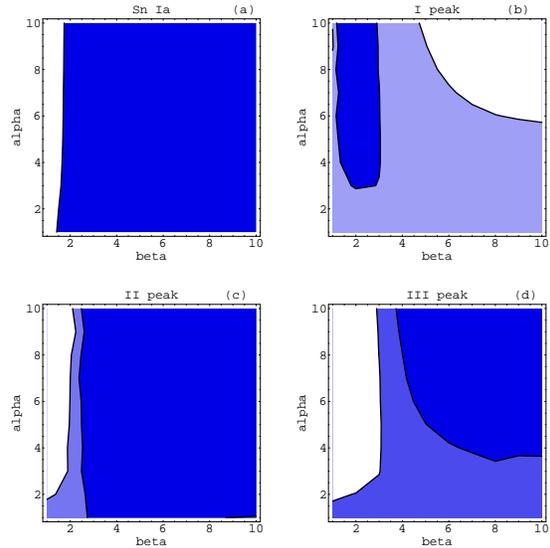}
\caption{ Likelihood contour plots  for SnIa, I, II and III acoustic peaks.
The blue region is the $68\%$ confidence region while the $90\%$ is
the light blue one. For the SnIa the white region correspond to $2\sigma$.
The position of the third CMB acoustic peak strongly constrains the acceptable parameter
space.
\label{fig3}}
\end{figure}

\subsection{Constraints from Doppler peaks and SnIa}

We now compute the position of the three Doppler peaks
$l_{1},l_{2}$ and $l_{3}$ using Eq.~(\ref{peak}). In addition
to the parameter space used in the supernovae analysis we consider 
the physical baryon density and the scalar spectral index varying respectively in
the range $\Omega_{b}h^2\in(0.018,0.026)$ and $n\in(0.9,1.1)$. 
The Hubble constant is set to $h=0.70$ in agreement with the recent 
HST observations (Freedman et al. 2000).
 The predicted peak multipoles in the CMB are then compared with
those mesured in the BOOMERANG  and DASI spectra (De Bernardis et al. 2001).
Note, that the third peak has been detected in the BOOMERANG data
but not in the DASI data.
Furthermore the authors of De Bernardis \textit{et al.} (2001), with a model independent analysis,
estimated
the position of the peaks accurately at $1\sigma$. However because the errors 
associated with the data are non Gaussian, to be conservative we take our $1\sigma$ errors on the data 
to be larger  than those
reported in De Bernardis \textit{et al.} (2001), so that our analysis is significant up to $2\sigma$. 
We then evaluate a gaussian likelihood function $\mathcal{L}^{Peaks}(\alpha,\beta
,\Omega_{Q},\Omega_{b}h^2,n)$.
The combined one-dimensional likelihood function for the peaks is shown in Fig.2b, where we find $\Omega_{Q}=0.69\pm
\genfrac{}{}{0pt}{}{0.13}{0.10}$. The likelihood for all the
data sets combined is shown in Fig.2c, where we find $\Omega_{Q}=0.75\pm
\genfrac{}{}{0pt}{}{0.09}{0.08}$. These results are in agreement
with the analysis of Efstathiou (2000), Netterfield \textit{et al.} (2001) and Baccigalupi \textit{et al.} (2001).

\begin{figure}[ht]
\plotone{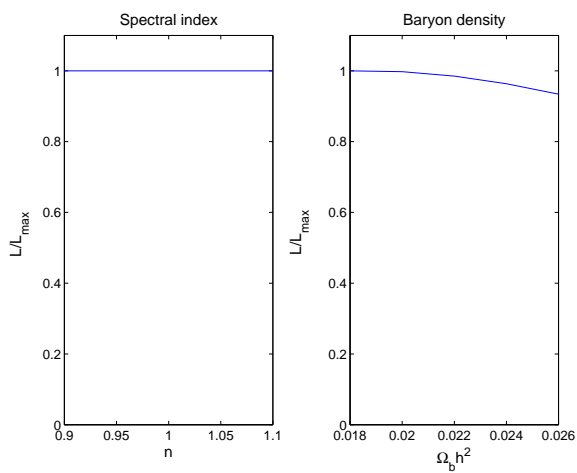}
\caption{ One-dimensional likelihood for $n$ and $\Omega_{b}h^2$.
\label{fig4}}
\end{figure}
\begin{figure}[ht]
\plotone{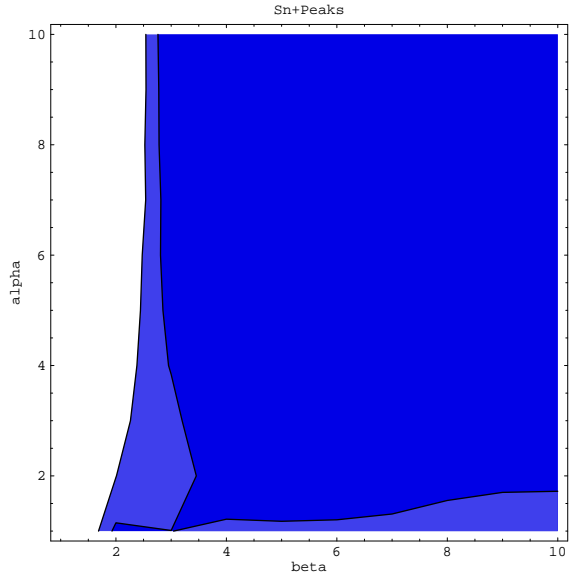}
\caption{ Two-dimensional likelihood for SnIa and CMB with $1$ (dark blue) and $2\sigma$ (light blue) contours.
\label{fig5}}
\end{figure}
\begin{figure}[htb]
\plotone{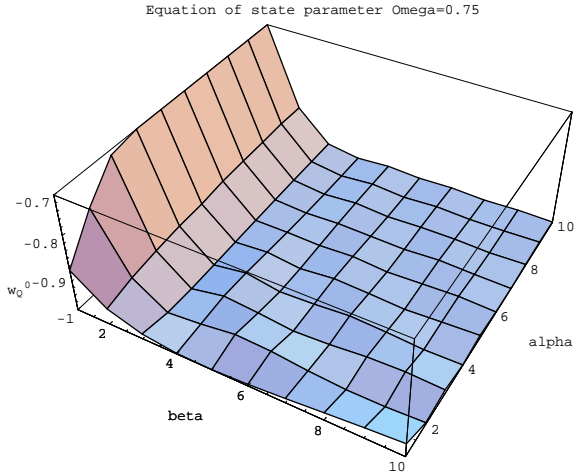}
\caption{Equation of state parameter against $\alpha$ and $\beta$ for 
$\Omega_{Q}=0.75$. The $2\sigma$ contours correspond to models with $w_{Q}^0\sim-1$.
\label{fig6}}
\end{figure}

The likelihood functions, combining all the data for the CMB peaks,
 for the scalar spectral index and the physical baryon density 
are shown in Fig.4. Since the dependence of the peak multipoles on $\Omega_{b}h^2$
and $n$ is small, it is not possible to obtain some significant constraints on these cosmological parameters
using the location of the Doppler
peaks. 
In Fig.3b-3d we plot the two-dimensional likelihood function in the plane $\alpha-\beta$ for each peak,
obtained after having 
marginalized over $\Omega_{Q}$, $\Omega_{b}h^2$ and $n$. 
Their shape reflects the accuracy in the estimation of the position of the peaks.
Actually the first one is very well resolved, while we are less confident with the location of the 
second and third peak. Therefore their likelihoods are more spread and flat in the $\alpha-\beta$ plane. 
The $1\sigma$ confidence 
contour (Fig.3b) for the first acoustic peak constrains the slopes of our potential in the range:
$3 \leq \alpha \leq 10$ and $1 \leq \beta \leq 3$. In particular the likelihood has a maximum at $\alpha=9$ and
$\beta=2$, corresponding to an equation of state $w_{Q}^0=-0.8$ for $\Omega_{Q}=0.7$, in agreement with the
recent analysis by Baccigalupi \textit{et al.} (2001).
However, the second and third peaks constrain a region where the equation of state is compatible with the cosmological constant value.
Therefore the effect of including all the data in the likelihood analysis is to move the constraint from
models with $w_{Q}^0\sim-0.8$ to models with an equation of state $w_{Q}^0\sim-1$.
As we can see in Fig.6 the values of $\alpha$ and $\beta$, allowed by the likelihood including all the 
data (Fig.5), correspond to our models with values of $w_{Q}^0$ in the range 
$-1\leq w_{Q}^0 \leq 0.93$ at $2\sigma$ for our prior probability $\Omega_{Q}=0.75$. The reason for
such a strong constraint is due to the assumed accurate determination of the third peak, in that it is insensitive to pre-recombination 
effects. In particular peak multipoles are
shifted toward larger values as $w_{Q}^0$ approaches the 
cosmological constant value. This is because, in models with $w_{Q}^0\sim-1$
the universe starts to accelerate earlier than in those with $w_{Q}^0>-1$, consequently the 
distance to the last scattering surface is further and hence the sound horizon at the 
decoupling is projected onto smaller angular scales. Since the location
of the third peak inferred by De Bernardis \textit{et al.} (2001) 
is at $l_{3}=845\pm
\genfrac{}{}{0pt}{}{12}{23}$, values of $w_{Q}^0\sim-1$ fit this
multipole better than models with $w_{Q}^0>-1$.
However we want to point out that at $1\sigma$ the position of the first peak is 
inconsistent with the position of the other two. A possible explanation of this discrepancy
is that the multipoles $l_{2}$ and $l_{3}$  are less sensitive to small shift 
induced by the dependence on $\Omega_{b}h^2$ and $n$. Therefore we can obtain a different constraint
on the dark energy equation of state if we consider the peaks individually.

\section{Conclusions}
The location of the sound horizon is very sensitive to the dark energy 
contribution.
Due to the strong degeneracy in the shape of the CMB spectrum, a certain 
class of Quintessence models can be better
constrained using only the acoustic peaks. We have applied a likelihood 
analysis to constrain the shape of the Quintessence
potential, 
based on both the supernovae type Ia data and the positions of the
CMB peaks. Assuming a flat space-time and making use only of the position of the 
first Doppler peak we find the best fit for models $w_{Q}^0\sim-0.8$ for
$\Omega_{Q}=0.7$ prior value. The combined analysis, including all three peaks and SnIa, gives the 
best fit for $\Omega_{Q}=0.75\pm
\genfrac{}{}{0pt}{}{0.08}{0.09}$. We have found in particular that the determination of third peak
in the BOOMERANG data limits the equation of state parameter at $2\sigma$ in the 
range $-1\leq w_{Q}^0\leq-0.93$ for $\Omega_{Q}$ with this prior value.
This has an important implication for minimally coupled 
Quintessence models.
Actually they must behave similarly to a
cosmological constant, therefore inverse power law is disfavoured. 
In fact, an equation of state parameter $w_{Q}^0\sim-1$ implies the Quintessence field is 
undergoing small damped oscillations around a minimum or evolving in a
very flat region of the potential. For these reasons models like 
the double exponential
potential (Barreiro et al. 2000) or the single modified exponential potential
 (Skordis \&\ Albrecht 2000) pass this constraint, even though they are not included
 in our analysis. Another important caveat
 is that this study does not 
take into account Quintessence
scenarios where the contribution of the dark energy density in radiation or 
early in
matter dominated eras is not negligible. In such a case we would have 
to take into account physical effects 
not only on distance
measurements, but also on the structure formation process itself.
These models and the non-minimally coupled ones therefore could yet be distinguished
 from a pure $\Lambda CDM$ model.
We still require a more complete analysis to 
understand the nature of the dark energy, but this paper points out
that it is possible to constrain certain classes of models far more than 
was previously realised.

\acknowledgments

We are very grateful to the referee for making a number of constructive comments and to 
L. Amendola, A. Liddle, N.J. Nunes and J. Weller for useful discussions and comments.

\end{document}